\input{aipcheck}

\documentclass[sort&compress
    ,final            
  ]
  {aipproc}

\layoutstyle{8x11single}

\usepackage[latin1]{inputenc}
\usepackage[T1]{fontenc}
\usepackage[english,american]{babel}
\usepackage{amstext,amsmath,amssymb,amsthm,mathrsfs}
\usepackage{epsfig}
\usepackage{graphicx} 
\usepackage[dvipsnames]{xcolor}

\begin{document}

\title{Neutrinoless Double Beta Decay and Neutrino Masses
\footnote{Poster presented at the International Workshop on Grand Unified Theories, Yukawa Institute for Theoretical Physics, Kyoto, Japan, March 2012.}}


\author{Michael Duerr}{
  address={Max-Planck-Institut f\"ur Kernphysik, Saupfercheckweg 1, 69117 Heidelberg, Germany}
}

\begin{abstract}
Neutrinoless double beta decay ($0\nu\beta\beta$) is a promising test for lepton number violating physics beyond the standard model (SM) of particle physics. There is a deep connection between this decay and the phenomenon of neutrino masses. In particular, we will discuss the relation between $0\nu\beta\beta$ and Majorana neutrino masses provided by the so-called Schechter--Valle theorem in a quantitative way. Furthermore, we will present an experimental cross check to discriminate $0\nu\beta\beta$ from unknown nuclear background using only one isotope, i.e., within one experiment.
\end{abstract}

\maketitle


\section{Introduction to neutrinoless double beta decay}
\begin{figure}
\centering
\begin{minipage}{.49\linewidth}
 \includegraphics[width=.6\linewidth]{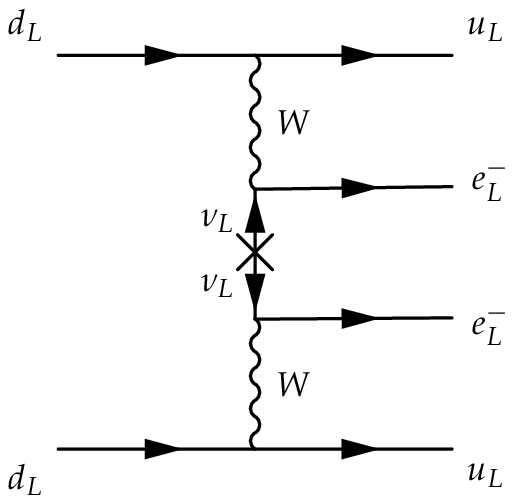}
\end{minipage}
\begin{minipage}{.49\linewidth}
\includegraphics[width=\linewidth]{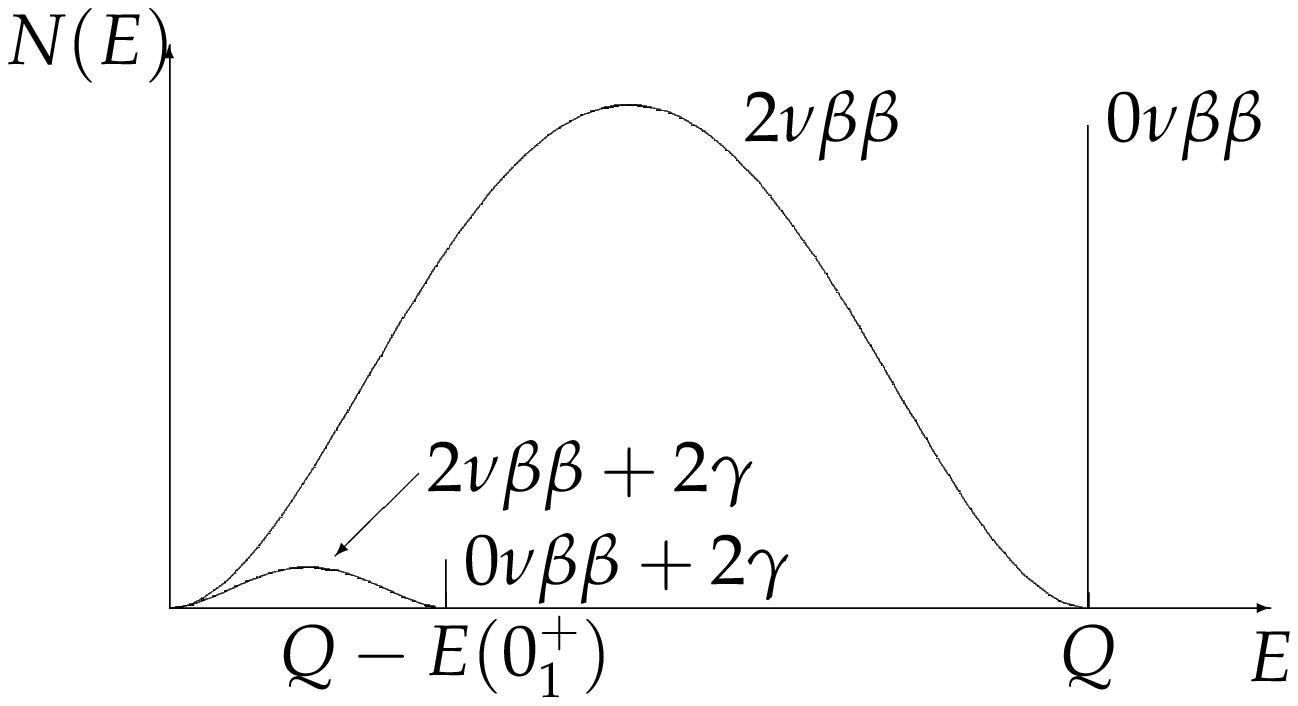}
\end{minipage}
\caption{
\textit{Left panel:} $0\nu\beta\beta$ mediated by light neutrino exchange, which is considered to be the standard mechanism. Of course, it is definitely not the only possible decay mechanism, as many models for physics beyond the standard model predict lepton number violation and therefore may trigger this decay via a different set of virtual particles. 
\textit{Right panel:} Sum energy spectrum of the emitted electrons for the two modes of $\beta\beta$ (not to scale). The curves labeled $0\nu\beta\beta$ and $2\nu\beta\beta$ are for the transition to the ground state of the daughter nucleus. Note that the energy spectrum for the two-neutrino mode is continuous because the emitted neutrinos may take away an arbitrary amount of energy, whereas the spectrum for the zero-neutrino mode is a single peak at the maximum energy $Q$ of the decay. The curves labeled $0\nu\beta\beta + 2\gamma$ and $2\nu\beta\beta + 2 \gamma$ are for the decay to the first excited $0^+$ state, which will also be discussed here. These lines lie at lower energies because energy is taken away by the emitted photons. Additionally, a lower number of decays to excited states than decays to the ground state occurs.}\label{fig:0nbbstandard}
\end{figure}

In some even-even nuclei single beta decay is energetically forbidden or strongly suppressed. In this case, double beta decay ($\beta\beta$) is the only allowed transition. It may occur in two modes:
\begin{equation}
(Z,A) \rightarrow (Z+2,A) + 2 e^- + 2 \bar{\nu}_e \quad (2\nu\beta\beta)
\end{equation}
and
\begin{equation}\label{eq:proc0nu}
(Z,A) \rightarrow (Z+2,A) + 2 e^-  \quad (0\nu\beta\beta) \, .
\end{equation}
$2\nu\beta\beta$ is allowed in the SM because it conserves lepton number. In contrast, $0\nu\beta\beta$ is only possible if lepton number is broken in Nature. Experimentally, the two decays may be distinguished by measuring the sum energy spectrum of the emitted electrons (see right panel of Fig.~\ref{fig:0nbbstandard}, curves labeled $0\nu\beta\beta$ and $2\nu\beta\beta$). $2\nu\beta\beta$ has been detected experimentally with half-lives of the order $10^{20}\, \mathrm{y}$, whereas $0\nu\beta\beta$ has not been seen experimentally yet; best bounds on the half-life of this decay are of the order $10^{25}\, \mathrm{y}$. 

The total decay rate $\Gamma^{0\nu}$ of $0\nu\beta\beta$ (if mediated by light neutrino exchange, see left panel of Fig.~\ref{fig:0nbbstandard}) and the corresponding half-life $T^{0\nu}_{1/2}$ are given by
\begin{equation}\label{eq:decay_rate}
 \Gamma^{0\nu} / \ln 2 =(T^{0\nu}_{1/2})^{-1} = \left|m_{ee}\right|^2 \left|\mathcal{M}^{0\nu}\right|^2 G^{0\nu}(Q,Z)\, ,
\end{equation}
where $\mathcal{M}^{0\nu}$ is the so-called nuclear matrix element encoding the nuclear physics involved and $G^{0\nu}(Q,Z)$ is an exactly calculable phase space factor. Given this particular decay mechanism, it is therefore possible to measure the so-called effective Majorana mass of the electron neutrino
\begin{equation}
 \left|m_{ee}\right| = \left| \sum_i U_{ei}^2 m_i \right| ,
\end{equation}
where $U_{ei}$ are the elements of the first line of the Pontecorvo--Maki--Nagawa--Sakata mixing matrix, and $m_i$ are the light neutrino mass eigenvalues. 

More details of neutrinoless double beta decay may be found in the recent reviews \cite{Bilenky:2010zz,Rodejohann:2011mu,Vergados:2012xy}.

In this contribution, I will shortly review the results from \cite{Duerr:2011yh,Duerr:2011zd}, both connected to the topics of neutrinoless double beta decay and neutrino masses.

\section{On the quantitative impact of the Schechter--Valle theorem}
\begin{figure}
 \centering
\includegraphics[scale=1]{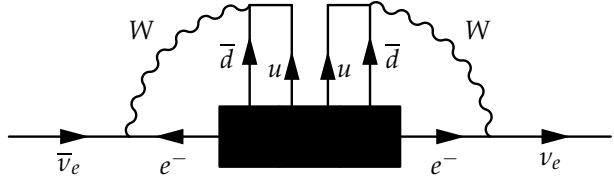} 
\caption{Majorana neutrino mass term generated by $0\nu\beta\beta$ \cite{Schechter:1981bd}.}\label{fig:blackbox}
 \end{figure}

The Schechter--Valle theorem (black box theorem) \cite{Schechter:1981bd} establishes the following relation: If $0\nu\beta\beta$ is seen, the neutrino is a Majorana particle (it has a Majorana mass term). The black box (see Fig.~\ref{fig:blackbox}) may contain any new physics mechanism triggering $0\nu\beta\beta$. We know that neutrino masses are tiny in relation to the charged lepton masses, but the mass is generated at four-loop order. So the question arises how big this mass is. In \cite{Duerr:2011zd}, a quantitative analysis of the mass correction induced by the Schechter--Valle theorem was performed for point-like operators (heavy new physics contributions). 

To perform the loop calculations, the most general Lorentz-invariant Lagrangian for $0\nu\beta\beta$ (point-like operators) is needed~\cite{Pas:2000vn} ($G_F$ is the Fermi coupling and $m_p$ is the proton mass):

\begin{equation}\label{eq:generallagrangian}
 \mathcal{L}=\frac{G_F^2}{2}m_p^{-1} \left( \epsilon_1 JJj + \epsilon_2 J^{\mu\nu} J_{\mu\nu} j + \epsilon_3 J^\mu J_\mu j + \epsilon_4 J^\mu J_{\mu\nu} j^\nu + \epsilon_5 J^\mu J j_\mu \right)\, ,
\end{equation}
where the hadronic currents are given by
\begin{equation}\label{eq:hadronic}
J = \overline{u}\left( 1 \pm \gamma_5 \right) d , \; J^\mu = \overline{u} \gamma^\mu \left( 1 \pm \gamma_5 \right) d, \;
 J^{\mu\nu} = \overline{u} \frac{i}{2} \left[\gamma^\mu, \gamma^\nu\right] \left( 1 \pm \gamma_5 \right) d \, ,
\end{equation}
and the leptonic currents are given by
\begin{equation}\label{eq:leptonic}
 j = \overline{e}\left( 1 \pm \gamma_5 \right) e^c , \; j^\mu = \overline{e} \gamma^\mu \left( 1 \pm \gamma_5 \right) e^c\, .
\end{equation}
From the experimental non-observation of neutrinoless double beta decay limits on the coupling constants $\epsilon_i$ can be deduced (on-axis evaluation). See \cite{Pas:2000vn} and \cite{Duerr:2011zd}, the latter for results using updated matrix element calculations. 

Plugging one operator at a time into the black box in Fig.~\ref{fig:blackbox} and performing the loop calculations, we can find a mass correction for all the possible operators giving neutrinoless double beta decay from Eq.~\ref{eq:generallagrangian}. More details of the calculations can be found in \cite{Duerr:2011zd}. As an example, here we give the result~\cite{Duerr:2011zd} for the operator $J^\mu_R J_{\mu R} j_L$ [indices $R$ and $L$ refer to a particular chirality of the operators in Eqs.~\ref{eq:hadronic} and \ref{eq:leptonic}, using the projectors $P_L = \frac{1}{2}(1-\gamma_5)$ and $P_R = \frac{1}{2}(1+\gamma_5)$]:
\begin{equation}
m_\nu = \delta m_\nu  \sim 9.4\times 10^{-25} \, \mathrm{eV}.
\end{equation}

We find that the radiatively generated masses are many orders of magnitude smaller than the observed neutrino masses.
Lepton number violating new physics (at tree-level not necessarily related to neutrino masses) may induce black box operators which explain an observed rate of $0\nu\beta\beta$.
The smallness of the black box contributions, however, implies that other neutrino mass terms (Dirac/Majorana) must exist.
If the $\nu$'s are mainly Majorana particles, the mass mechanism is the dominant part of the black box operator.
If the $\nu$'s are mainly Dirac particles, other lepton number violating new physics dominates $0\nu\beta\beta$. Translating an observed rate of $0\nu\beta\beta$ into neutrino masses would then be completely misleading. 

\section{Consistency test of $\boldsymbol{0}\boldsymbol{\nu}\boldsymbol{\beta}\boldsymbol{\beta}$ with one isotope}

\begin{figure}
\centering
\begin{minipage}{.49\linewidth}
 \includegraphics[width=\linewidth]{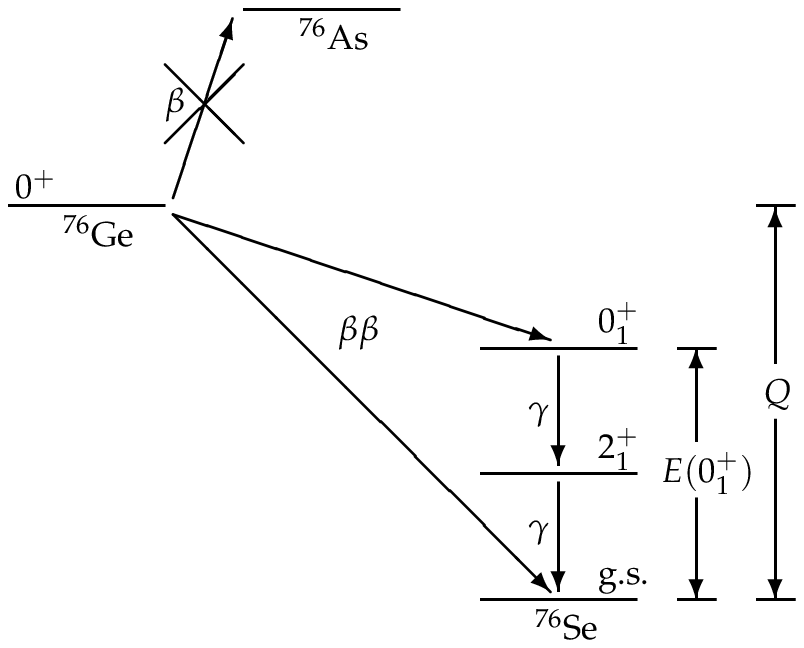}
\end{minipage}
\begin{minipage}{.49\linewidth}
\includegraphics[width=\linewidth]{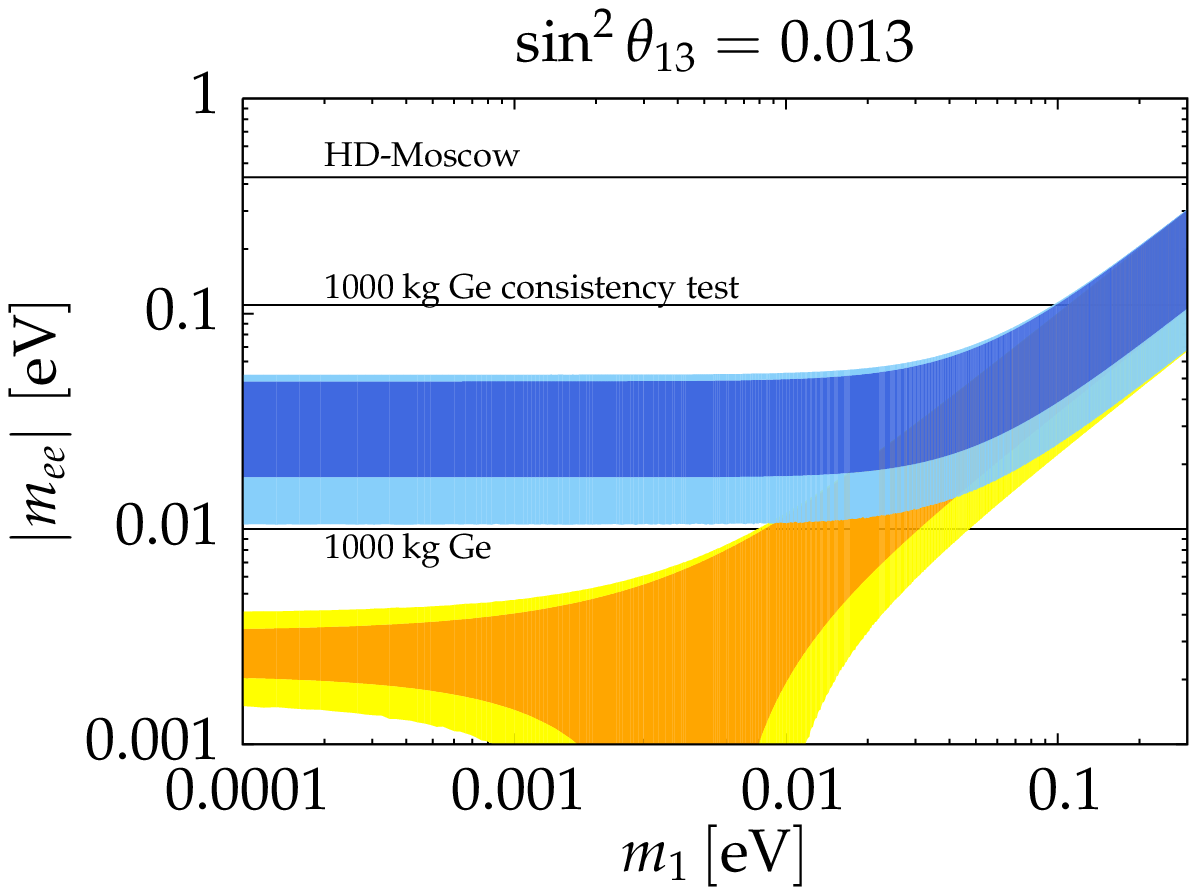}
\end{minipage}
\caption{
\textit{Left panel:} Level scheme of $^{76}$Ge and its daughter nuclei. Standard beta decay is energetically forbidden. Besides the usual decay to the ground state (g.s.), we will discuss the decay to the first excited $0^+$ state (labeled $0_1^+$). 
\textit{Right panel:} The effective Majorana neutrino mass $\left| m_{ee} \right|$ as a function of the lightest neutrino mass eigenvalue for inverted (upper band) and normal (lower band) hierarchy. Both hierarchies overlap in the quasi degenerate case. Bold colors denote the best fit values range varying the CP phases, light colors give the corresponding $3\sigma$ ranges. The best fit value from the Heidelberg--Moscow experiment \cite{KlapdorKleingrothaus:2004wj} is marked. A future 1 ton $^{76}$Ge experiment \cite{Schonert:2005zn} could probe the inverted hierarchy down to $\left| m_{ee}\right|=0.01 \, \mathrm{eV}$. At the same time, it could be used for the proposed consistency test covering the quasi degenerate region.}\label{fig:0nbbexcited}
\end{figure}

Usually, only the decay to the ground state of the daughter nucleus is considered in $0\nu\beta\beta$ experiments. However, all $0\nu\beta\beta$ isotopes have an excited $0^+$ state accessible by this decay, see Fig.~\ref{fig:0nbbexcited} (left panel) for more details.
This state may be used for a consistency test \cite{Duerr:2011yh} which allows to discriminate unknown nuclear background lines from $0\nu\beta\beta$ using only one isotope.

The ratio of decay rates to the two different states of the daughter nucleus is given by 
\begin{equation}
 \frac{\Gamma_{0_1^+}}{\Gamma_{\mathrm{g.s.}}}= \frac{(Q-E(0_1^+))^n}{Q^n} \times \left(\frac{\mathcal{M}^{0_1^+}}{\mathcal{M}^{\mathrm{g.s.}}}\right)^2 \, .
\end{equation}
Here, $n=5$ for $0\nu\beta\beta$ and $n=11$ for $2\nu\beta\beta$. $\mathcal{M}^{\mathrm{g.s.}}$ and $\mathcal{M}^{0_1^+}$ are the matrix elements for the decay to the ground state and the first excited $0^+$ state, respectively. $Q$ and $E(0_1^+)$ are defined in Fig.~\ref{fig:0nbbexcited}. Due to the lower energy difference, the rate of decays to excited states is lower than the rate of decays to the ground state. The ratio is given for the two most promising isotopes in Table~\ref{tab:excitedstates}, a full list can be found in \cite{Duerr:2011yh}. 

The question arises if the excited states can be used as a cross check for $0\nu\beta\beta$, this means to prove this decay within one experiment using only one isotope in two ways. Such a test could be desirable in future large-scale experiments due to the high costs. Further experimental considerations and a discussion of possible backgrounds to be taken into account may be found in \cite{Duerr:2011yh}.

The result is depicted in Fig.~\ref{fig:0nbbexcited}, right panel. For ton-scale detectors, the method works down to $\left|m_{ee} \right| = 100\,\mathrm{meV}$. A large scale detector could therefore simultaneously provide a consistency test for a certain range of Majorana masses, while it would also be sensitive to lower values of the effective Majorana mass $\left| m_{ee}\right|$. 

\renewcommand{\arraystretch}{1.5}
\begin{table}
\centering
\begin{tabular}{cccccc}
\hline
Decay mode & $Q \; [\mathrm{keV}]$ & $E (0_1^+) \; [\mathrm{keV}]$ & $\mathcal{M}_{0\nu}^{\mathrm{g.s.}}$& $\mathcal{M}_{0\nu}^{0_1^+}$ & $\Gamma_{0_1^+} / \Gamma_{\mathrm{g.s.}}$ \\
 \hline \hline
$_{32}^{76}$Ge$\rightarrow _{34}^{76}$Se & $2039.04 \pm 0.16$ \cite{Rahaman:2007ng} & 1122 & 5.465 & 2.479 & $3.79 \times 10^{-3}$\\
$_{~60}^{150}$Nd$\rightarrow _{~62}^{150}$Sm & $3371.38 \pm 0.20$ \cite{Kolhinen:2010zz} &  740 & 2.321 & 0.395 &  $8.39 \times 10^{-3}$\\
\hline 
\end{tabular}
\caption{The two most promising isotopes for the proposed cross check and the corresponding parameters. Excerpt from \cite{Duerr:2011yh}. The energy $E(0_1^+)$ of the first excited states is taken from \cite{toi:1998}. The given nuclear matrix elements are obtained within the IBM-2 model \cite{Barea:2009zza,iac2010}.}\label{tab:excitedstates}
\end{table}

\vspace*{-1ex}

\section{Summary}

We discussed two different topics related to neutrinoless double beta decay and neutrino masses. 

In the first part, the relation between the effective operators responsible for neutrinoless double beta decay and neutrino mass operators was discussed. We found that one has to be careful relating a possible observation of neutrinoless double beta decay fully to neutrino masses, while a tiny Majorana contribution via the well-known black box diagram is guaranteed. The possible implications were discussed.

In the second part, we showed that it is possible to check within a single experiment whether a possibly observed signal is indeed $0\nu\beta\beta$ or due to some unknown nuclear background. Usually, it is proposed that a second experiment using a different isotope should be used to provide this test. We showed that effort could be combined into on large detector instead of building several ones. 


\vspace*{-1ex}

\begin{theacknowledgments}
I would like to thank the organizers of the GUT2012 workshop at the Yukawa Institute for Theoretical Physics in Kyoto, Japan, for the nice workshop and the possibility to present my poster. Moreover, I would like to thank my collaborators Manfred Lindner, Alexander Merle, and Kai Zuber, together with whom the work presented here was performed. I am supported by the International Max Planck Research School (IMPRS) ``Precision Tests of Fundamental Symmetries'' of the Max Planck Society. 
\end{theacknowledgments}



\bibliographystyle{aipproc}   


\vspace*{-1ex}

\bibliography{bib_duerr}

\end{document}